\documentclass[journal,final]{IEEEtran}
\IEEEoverridecommandlockouts                              
                                                          
\usepackage{cite}
\usepackage{amsmath,amssymb,amsfonts, amsthm}
\usepackage{algorithm} 
\usepackage{graphicx}
\graphicspath{{fig/}{../fig/}}
\usepackage{subfigure}
\usepackage{amssymb}
\usepackage{cuted}
\usepackage{widetext}
\usepackage{multicol}
\usepackage{lipsum}  
\usepackage{cite}
\usepackage{textcomp}
\def\BibTeX{{\rm B\kern-.05em{\sc i\kern-.025em b}\kern-.08em
    T\kern-.1667em\lower.7ex\hbox{E}\kern-.125emX}}
    
\usepackage{color}

\providecommand{\com[2]}{\begin{tt}[#1: #2]\end{tt}}

\newcommand{\bydef}{\stackrel{\Delta}{=}}

\newcommand{\beq}{\begin{equation}}
\newcommand{\eeq}{\end{equation}}
\newcommand{\beqa}{\begin{eqnarray}}
\newcommand{\eeqa}{\end{eqnarray}}
\newcommand{\beqan}{\begin{eqnarray*}}
\newcommand{\eeqan}{\end{eqnarray*}}
\newcommand{\bef}{\begin{figure}}
\newcommand{\enf}{\end{figure}}

\definecolor{magenta}{cmyk}{0.5, 1, 0, 0}
\definecolor{greenedu}{cmyk}{1, 0, 1, 0.1}
\definecolor{cyanedu}{cmyk}{1, 0, 0, 0.1}

\newcommand{\bi}{\begin{itemize}}
\newcommand{\ei}{\end{itemize}}
\newcommand{\bc}{\begin{center}}
\newcommand{\ec}{\end{center}}
\newcommand{\ba}{\begin{array}}
\newcommand{\ea}{\end{array}}
\newcommand{\be}{\begin{equation}}
\newcommand{\ee}{\end{equation}}
\newcommand{\beno}{\begin{equation*}}
\newcommand{\eeno}{\end{equation*}}
\newcommand{\beqna}{\begin{eqnarray}}
\newcommand{\eeqna}{\end{eqnarray}}
\newcommand{\bd}{\begin{displaymath}}
\newcommand{\ed}{\end{displaymath}}
\newcommand{\beqnd}{\begin{eqnarray*}}
\newcommand{\eeqnd}{\end{eqnarray*}}

\renewcommand{\ni}{\noindent}

\newtheorem{theorem}{\bf Theorem}[section]
\newtheorem{lemma}{\bf Lemma}[section]

\newtheorem{proposition}{\bf Proposition}[section]

\newtheorem{definition}{\bf Definition}

\usepackage{color}

\definecolor{red}{rgb}{1,0,0}

\definecolor{blu}{rgb}{0,0,1}

\definecolor{gre}{rgb}{0,0.7,0.3}

\definecolor{bz}{rgb}{1,.7,0}

\include{temp_jh_ref.bib}

\renewcommand\qedsymbol{$\blacksquare$}
\makeatletter
\renewenvironment{proof}[1][\proofname]{\par
  \normalfont \topsep6\p@\@plus6\p@\relax
  \trivlist
  \item[\hskip\labelsep
        \itshape
    #1\@addpunct{.}]\ignorespaces
}{%
  \nolinebreak\qedsymbol\endtrivlist\@endpefalse
}
\makeatother

\begin{document}

 \title{Identifiability of dynamic networks: the essential r\^{o}le of dources and dinks}

\author{Eduardo Mapurunga, Michel Gevers, \IEEEmembership{Life Fellow, IEEE,} and Alexandre S. Bazanella, \IEEEmembership{Senior Member, IEEE} 
\thanks{Eduardo Mapurunga and Alexandre S.  Bazanella are with the Data Driven Control Group, Department of Automation and Energy, 
Universidade Federal do Rio Grande do Sul, Porto Alegre-RS, Brazil, \{eduardo.mapurunga, bazanella\}@ufrgs.br}
\thanks{Michel Gevers is with the Institute of Information and Communication Technologies, Electronics and Applied Mathematics (ICTEAM), UCLouvain, Louvain la Neuve, Belgium,
michel.gevers@uclouvain.be.}
\thanks{This work  was supported in part by the Coordena{\c c}{\~ a}o de Aperfei{\c c}oamento de Pessoal de N{\'i}vel Superior - Brasil (CAPES) - Finance Code 001, by Conselho Nacional de Desenvolvimento Cient{\' i}fico e Tecnol{\' o}gico (CNPq),
by Wallonie-Bruxelles International (WBI), by a WBI.World excellence fellowship, and by a Concerted Research Action (ARC) of the French Community of Belgium.}
}


\maketitle




\begin{abstract}
The paper \cite{bazanella-gevers-hendrickx-network-2019} presented the first results on generic identifiability of dynamic networks with partial excitation and partial measurements, i.e. networks where not all nodes are excited or not all nodes are measured. One key contribution of that paper was to establish a set of necessary conditions on the excitation and measurement pattern (EMP) that guarantee generic identifiability. In a nutshell, these conditions established that all sources must be excited and all sinks measured, and that all other nodes must be either excited or measured. In the present paper, we show that two other types of nodes, which are defined by the local topology of the network, play an essential r\^{o}le in the search for a valid EMP, i.e. one that guarantees generic identifiability. We have called these nodes dources and dinks. We show that a network is generically identifiable only if, in addition to the above mentioned conditions, all dources are excited and all dinks are measured. We also show that sources and dources are the only nodes in a network that always need to be excited, and that sinks and dinks are the only nodes that need to be measured for an EMP to be valid.
\end{abstract}


\section{Introduction}\label{intro}


This work deals with identifiability of dynamic networks, which has been an active research topic in the control community over the last decade. The network framework used in this paper was introduced in \cite{vandenhof-dankers-heuberger-etal-identification-2013}, where signals are represented as nodes of the network which are related to other nodes through transfer functions.  To such dynamic network, one can associate a directed graph, where the transfer functions, also called modules, are the edges of the graph and the node signals are its vertices.

In \cite{vandenhof-dankers-heuberger-etal-identification-2013}, it was assumed that all nodes are excited and measured. As a result, an input-output matrix of the network,  denoted $T(z)$, can be defined, which can always be identified from these excitation and measurement data. The network identifiability question is then whether the network matrix, denoted $G(z)$  (whose elements are the transfer functions relating the nodes) can be recovered from this input-output transfer matrix $T(z)$. 
In subsequent works, a range of new objectives were defined, from the  identification of the whole network  to identification of some specific part of the network 
\cite{gevers-bazanella-parraga-identifiability-2017, bazanella-gevers-hendrickx-etal-identifiability-2017, shi-cheng-vandenhof-generic-2022 , hendrickx-gevers-bazanella-identifiability-2019, everitt-bottegal-hjalmarsson-empirical-2018, gevers-bazanella-dasilva-practical-2018, van_waarde_necessary_2019, jahandari-materassi-sufficient-2021}. 
As for the assumptions on the signals, up to 2019, all contributions  assumed that either all nodes are excited, or all nodes are measured.  A typical question would be: given that all nodes are excited, which nodes must be measured in order to identify the whole network? 

The first identifiability results for networks with partial excitation and measurement were presented in \cite{bazanella-gevers-hendrickx-network-2019}. Thus, in that paper, the assumption that either all nodes are measured, or all nodes are excited, was removed. The paper first provided a necessary condition for identifiability of any network: each node must be either excited or measured, at least one node must be excited, and at least one node measured.  The paper \cite{bazanella-gevers-hendrickx-network-2019} also  presented identifiability conditions for two special classes of networks, namely trees and loops.

The results of  \cite{bazanella-gevers-hendrickx-network-2019}  inspired a new way of looking at the network identifiability problem. The question now becomes: what is a combination of excited nodes and measured nodes that allows the identifiability of the network? This led naturally to the definition of an  excitation and measurement pattern (EMP), namely the combination of excited nodes and measured nodes.    The concept of EMP was introduced in \cite{mapurunga-optimal-2021} where an EMP was called {\it valid} if it guarantees  the identifiability of the whole network. An EMP  was called {\it minimal} if it  guarantees the identifiability of the network using the smallest possible combination of excited and measured nodes \cite{mapurunga-optimal-2021}. This number is the {\it cardinality of the EMP.} Achieving identifiability of a network with a minimal EMP is of both theoretical and practical interest, since the excitation of a node or its measurement has a cost. On the other hand, having some flexibility in the choice of a valid EMP is also of practical interest, since these costs may be significantly different for different nodes. 
In evaluating the choice of an EMP for the identification of a network, one must of course remember that each node must be either excited or measured or both, as shown in \cite{bazanella-gevers-hendrickx-network-2019}. As a result, the cardinality of a valid EMP is always at least equal to $n$, the number of nodes.

The search for valid, and possibly minimal, EMPs began by looking at special structures. In \cite{bazanella-gevers-hendrickx-network-2019} a necessary and sufficient condition was given for the identifiability of a tree; it showed that a tree can possibly be identified with an EMP of cardinality $n$.  
In \cite{mapurunga-identifiability-2021} necessary and sufficient conditions were derived for the identifiability of some classes of parallel networks. 
In \cite{mapurunga-gevers-bazanella-necessary-2022}  necessary and sufficient conditions were given for the identifiability of loops. It was shown that any loop with more than 3 nodes can  be identified with a minimal EMP of cardinality $n$, and that constructing EMPs for loops - even minimal EMPs -  is very easy. A novel  approach to the generic identifiability of a network with partial excitation and measurement was developed in  \cite{legat-hendrickx-local-2020}, where the authors developed a local identifiability analysis which allows them to determine which transfer functions are  identifiable with probability one.

In \cite{mapurunga-gevers-baza-CDC2022}, we generalized the results derived in  \cite{bazanella-gevers-hendrickx-network-2019} for the identification of trees to a much wider class of networks, namely those that have the structure of a Directed Acyclic Graph (DAG), i.e.  a directed  graph that has no cycles. That paper focused on the construction of valid EMPs for the identification of a DAG. In the process of deriving necessary conditions for the identifiability of a DAG, we came up with a completely unexpected result. Whereas it has  been known for a long time that the identification of any network requires that all sources must be excited and all sinks must be measured, we showed that, besides sources and sinks, two other types of nodes play a particular r\^{o}le in the construction of a valid EMP for DAGs. We called them {\bf dources} and {\bf dinks}\footnote{See section \ref{sec:defs} for their definition.}, and we showed that identifiability of a DAG requires that all its dources be excited and all its dinks be measured.

To our surprise, we have since realized that the same holds true for the identifiability of any dynamic network. Thus, dources and dinks play an essential r\^{o}le in the search for a valid EMP, i.e. for the identifiability of a dynamic network. This is the main message of this paper, which is organized as follows. In Section~\ref{sec:defs} we define the dynamic networks that this paper deals with, we recall the definition of generic identifiability of a network, as well as the existing conditions on excitation and measurement for networks, and we introduce the new concept of dources and dinks. In Section~\ref{sec:necessary}, we present the first main new result of this paper: we show that, in addition to the previously known conditions for identifiability of a network, a network is generically identifiable only if all its dources are excited and all its dinks are measured. The results of Section~\ref{sec:ifandonlyif} go one step further. They show that sources and dources are the only nodes in a network that always require excitation, and that sinks and dinks are the only nodes that need to be measured for an EMP to be valid. The essential r\^{o}le of dources and dinks for generic identifiability of a dynamic network may appear bizarre at first. In Section~\ref{sec:understanding} we illustrate with an example the intuition that underpins these two new concepts. We also illustrate that, unless a node is a source or a dource, one can always identify the network without exciting that node. Finally, we conclude in Section~\ref{sec:conclu}.

\section{Definitions, Notations and Preliminaries}\label{sec:defs}

In this section, we introduce the dynamic networks that we will be dealing with, and we introduce the notations used for these networks and their associated graphs. We recall the necessary conditions for identifiability that were derived in \cite{bazanella-gevers-hendrickx-network-2019}, and we introduce the concept of  a valid Excitation and Measurement Pattern, namely a choice of excited nodes and measured nodes that guarantees identifiability. 

We consider dynamic networks composed of $n$ nodes (or vertices) which represent internal scalar signals $\left\lbrace w_k(t) \right\rbrace$ for $k \in  \{1, 2, \dots, n\}$.
These nodes are interconnected by discrete time transfer functions, represented by edges, which are entries of a \emph{network matrix} $G(z)$.
The dynamics of the network is given by the following equations:
\begin{subequations}
	\begin{align}
	w(t) &= G(z)w(t) + Br(t), \label{eq:dynet1} \\
	y(t) &= Cw(t), \label{eq:dynet2}
	\end{align}
\end{subequations}
where $w(t) \in \mathbb{R}^n$ is the node vector, $r(t) \in \mathbb{R}^m$ is the input vector, and $y(t) \in \mathbb{R}^p$ is the set of measured nodes, considered as the output vector of the network.
The matrix $B \in \mathbb{Z}_2^{n \times m}$, where $\mathbb{Z}_2 \triangleq \{0, 1\}$, 
is a binary selection matrix with a single $1$ and $n-1$ zeros  in each column; it selects the inputs affecting the nodes of the network.
Similarly, $C \in \mathbb{Z}_2^{p \times n}$ is a matrix with a single $1$ and $n-1$ zeros in each row that selects which nodes are measured. 

 To each network matirix $G(z)$ we can associate a directed graph $\mathcal{G}$ defined by the tuple $(\mathcal{V}, \mathcal{E})$, where $\mathcal{V}$ is the set of vertices and $\mathcal{E} \subseteq \mathcal{V} \times \mathcal{V}$ is the set of edges. The graph $\mathcal{G}$ defines the topology of the network.
A particular transfer function $G_{ij}(z)$ of the network matrix is called an incoming edge of node $i$ and outgoing edge of node $j$.
				Furthermore, for this transfer function, 
				we say that node $i$ is an out-neighbor of node $j$, and that node $j$ is an in-neighbor of node $i$.
				A node $j$ is \emph{connected} to node $i$ if there exists a directed edge from node $j$ to node $i$.
For the graph $\mathcal{G}$ associated to the network matrix $G(z)$ we introduce the following notations.
\begin{itemize}
	\item $\mathcal{W}$ - the set of all $n$ nodes;
	\item $\mathcal{B}$ - the set of excited nodes, defined by $B$ in (\ref{eq:dynet1});
	\item $\mathcal{C}$ - the set of measured nodes, defined by $C$ in (\ref{eq:dynet2});
	\item $\mathcal{F}$ - the set of sources: nodes with no incoming edges;
	\item $\mathcal{S}$ - the set of sinks: nodes with no outgoing edges;	
	\item $\mathcal{I}$ - the set of internal nodes, i.e. nodes that are neither a source nor a sink: $\mathcal{I} \triangleq \mathcal{W}\backslash(\mathcal{F}\cup\mathcal{S})$;
	\item $\mathcal{N}_j^-$ -  the set of in-neighbors  of node $j$;
	\item $\mathcal{N}_j^+$ - the set of out-neighbors of node $j$.
\end{itemize}
Additionally, we introduce the following two types of nodes discussed in the Introduction.
\begin{definition}
				A node $j$ is called a \textbf{dource} if all its in-neighbors have a directed edge to at least one of its out-neighbors.
				\label{def:dource}
\end{definition}

\begin{definition}
				A node $j$ is called a \textbf{dink} if it has at least one in-neighbor that has a directed edge to all  its
  out-neighbors.		\label{def:dink}
\end{definition}
We observe that a node can be both a dource and a dink. The attached figure illustrates an 8-node network in which 
node 2 is a dource, node 4 is a dink, and node 6 is both a dource and a dink. 

\begin{figure}[h!]
				\centering
				\includegraphics[width=0.95\columnwidth]{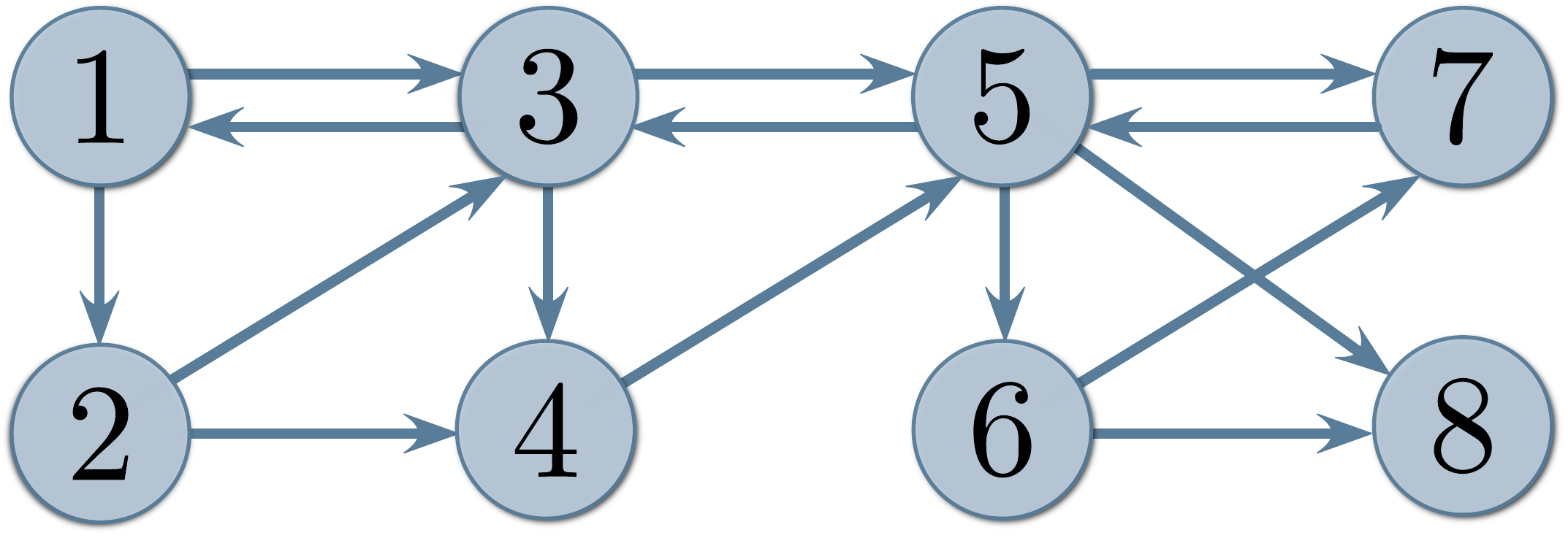}
				\caption{\normalsize A network with dources and dinks}
				\label{DD2DExample4}
\end{figure}

{\ni \bf Assumptions on the network matrix $G(z)$}\\
Throughout the paper, we shall make the following assumptions on the network matrix:
\begin{itemize}
	\item the diagonal elements are zero and all other elements are  proper;
	\item the network is well-posed in the sense that all minors of $(I - G(\infty))^{-1}$ are nonzero;
	\item $(I - G(z))^{-1}$ is proper and all its elements are stable.
\end{itemize}

One can represent the dynamic network in (\ref{eq:dynet1})-(\ref{eq:dynet2}) as an input-output model as follows
\begin{align}
	y(t) = M(z) r(t),\; \text{with}\; M(z) \triangleq C T(z) B.
	\label{eq:IOdynet}
\end{align}
where
\begin{align}
	T(z) \triangleq (I - G(z))^{-1}. \label{eq:Tdef}
\end{align}
Observe that the matrix $T(z)$ is nonsingular by construction. 

In analyzing the identifiability of the network matrix, it is assumed that the input-output model $M(z)$ is known; the identification of $M(z)$ from  input-output (IO) data $\lbrace y(t), r(t) \rbrace$ is a standard  identification problem, provided the input signal $r(t)$ is sufficiently rich. The question of identifiability of the network is then whether the network matrix $G(z)$ can be fully recovered from the  transfer matrix $M(z)$. We now give a formal definition of generic identifiability of the network matrix from the data $\lbrace y(t), r(t) \rbrace$ and from the graph structure.

\begin{definition}
(\cite{hendrickx-gevers-bazanella-identifiability-2019}) The network matrix $G(z)$ is generically identifiable from excitation signals applied to $\mathcal{B}$ and measurements made at $\mathcal{C}$ if, for any rational transfer matrix parametrization $G(P, z)$ consistent with the directed graph associated with $G(z)$, there holds
	\[
					C [I - G(P, z)]^{-1} B = C[I - \tilde{G}(z)]^{-1} B \implies G(P, z) = \tilde{G}(z),
	\]
	for all parameters $P$ except possibly those lying on a zero measure set in $\mathbb{R}^{N}$, where $\tilde{G}(z)$ is any network matrix consistent with the graph.
\end{definition}

In this paper, we discuss generic 
identifiability  in terms of which nodes must be excited and/or measured in order to guarantee identifiability of the network. Thus, we do not assume that either all nodes are excited or all nodes are measured, which was the common assumption until the publication of \cite{bazanella-gevers-hendrickx-network-2019}. Instead, we examine which constraints must be imposed on an EMP in order for it to be valid. Our results will expand on the 
following Proposition, which gives a necessary condition for generic identifiability of a network; it combines Theorem III.1 and Corollary III.1 of \cite{bazanella-gevers-hendrickx-network-2019}.
\begin{proposition}\label{corol1}
  The network matrix $G(z)$ is generically identifiable only if $\mathcal{B}, \mathcal{C} \neq \emptyset$, $\mathcal{F} \subset \mathcal{B}$, $\mathcal{S} \subset \mathcal{C}$  and $\mathcal{B} \cup \mathcal{C} = \mathcal{W}$.\label{cor:fromBandC}
\end{proposition}
\vspace{2mm}

Finding conditions that guarantee generic  identifiability of a given network  is equivalent to constructing an Excitation and Measurement Pattern (EMP)  that guarantees identifiability; such EMP is then called a {\emph valid EMP}.
The concept of EMP and of valid EMP, which led  to the concept of minimal EMP, was introduced in \cite{mapurunga-optimal-2021}. They are defined in the following. 
\begin{definition}
    A pair of selection matrices $B$ and $C$, with its corresponding pair of node sets $\mathcal{B}$ and $\mathcal{C}$,  is called an \textbf{excitation and measurement pattern} - EMP for short.
  An EMP is said to be \textbf{valid} if it is such that the network (\ref{eq:dynet1})-(\ref{eq:dynet2}) is generically identifiable.
  Let $\nu = |\mathcal{B}| + |\mathcal{C}|$ \footnote{$|\cdot|$ - Denotes the cardinality of a set.} be the cardinality of an EMP.
  A given EMP is said to be  \textbf{minimal} if it is valid and there is no other valid EMP with smaller cardinality.
  \label{def:EMP}
\end{definition}
The following result establishes a lower and an upper bound for the cardinality of a valid EMP for any network.
\begin{lemma}\label{cardinality}
The cardinality of a valid EMP for the identification of a dynamic network with $n$ nodes is at least equal to $n$ and at most equal to $2n-f-s$, where $f$ is the number of sources and $s$ the number of sinks.
\end{lemma}
\begin{proof} The lower bound results from Proposition~\ref{corol1}; it can actually be achieved for trees and loops  as shown in \cite{bazanella-gevers-hendrickx-network-2019,mapurunga-gevers-bazanella-necessary-2022}.
As for the upper bound, we know by Proposition~\ref{corol1} that all sources must be excited and all sinks measured, while the remaining $n-f-s$ nodes must be excited or measured. Assuming that these are all excited and measured, then the cardinality of the EMP is $f+s+ 2(n-f-s)= 2n-f-s$.
 \end{proof}
Proposition~\ref{corol1} showed that for an EMP to be valid, it must contain at least one excitation and one measurement, all sources must be excited and all sinks must be measured, and every other node must be either excited or measured. The objective of the present paper is to extend these results by showing that constraints must also be imposed on dources and dinks for an EMP to be valid. In particular, we shall show that, for an EMP to be valid, all dources must be excited (just like all sources) and all dinks must be measured (just like sinks). Beyond this extension of the earlier necessary conditions of Proposition~\ref{corol1}, we shall also present necessary and sufficient conditions on dources and dinks that any valid EMP must satisfy.

From now on, we drop the arguments $z$ and $t$ used in (\ref{eq:dynet1})-(\ref{eq:dynet2}) whenever there is no risk of confusion. Before we move on to the main results of this paper, we present a preliminary technical Lemma that will be useful in our further derivations.
\begin{lemma}
    Consider a dynamic network with network matrix $G$ and transfer matrix $T = (I - G)^{-1}$.
    The following relationships hold
    \begin{align}
						T_{ii} = 1 + \sum_{j = 1}^n T_{i j} G_{j i}, \label{eq:recursion1} \\ 
						T_{ik} =  \sum_{j = 1}^n T_{i j} G_{j k}, ~ \text{for} ~ k \neq i \label{eq:recursion2} \\
						\sum_{j = 1}^n G_{i j} T_{j k} = \sum_{j = 1}^n T_{i j} G_{j k}. \label{eq:recumirror}
    \end{align}
		\label{lem:recur}
\end{lemma}
\begin{proof}
    The proof follows from
    \begin{align}
            T (I - G) = I \iff
            T = I + TG
    .\end{align}
     The $i$-th row of $T$, denoted $T_i$, can be written as $T_i = I_i + T_i G$,  
		from which (\ref{eq:recursion1}) and (\ref{eq:recursion2}) follow. 
		As for   (\ref{eq:recumirror}), it follows from the fact that $(I - G) T = I = T( I - G) \iff  GT = TG$. 
\end{proof}

\section{A necessary condition on dources and dinks}\label{sec:necessary}

Our main result in this section is the following theorem. It presents a new set of necessary conditions for the identifiability of a network, which is an improvement on Proposition~\ref{corol1}.

\begin{theorem}
			A dynamic network is generically identifiable only if 
			\begin{enumerate}
							\item At least one node is excited and one node is measured;
							\item All sources and dources are excited;
							\item All sinks and dinks are measured;
							\item  All other nodes are either excited or measured.
			\end{enumerate}
			\label{theo:dourcesanddinks}
\end{theorem}
\begin{proof}
				All conditions except the excitation of dources and the measurement of dinks have been proved in Proposition~\ref{corol1}. We thus prove that dources must be excited; the proof for dinks follows by duality.

Let $D$ be a dource (see Definition \ref{def:dource}), and let $O$ be an out-neighbor of $D$ to which all its $k$ in-neighbors $I$ are  connected. 
Both $D$ and $O$ have dimension 1. 
				Let the remaining $n - 2 - k$ nodes of the network be labeled as $S$. The network matrix $G$ can then be partitioned as follows
								\begin{align}
								G = 
								\begin{bmatrix}
												0       & G_{D O} & G_{D I} & G_{D S} \\
												G_{O D} & G_{O O} & G_{O I} & G_{O S} \\
												G_{I D} & G_{I O} & G_{I I} & G_{I S} \\
												G_{S D} & G_{S O} & G_{S I} & G_{S S} 
								\end{bmatrix} 
								\label{eq:Gpartition1}
				.\end{align}
				
				Notice that from the definition of dource we have  $G_{DO} = 0$\footnote{Otherwise $O$ would also be an in-neighbor of $D$, and since all in-neighbors of $D$ must be connected to $O$, this would create a self-loop, which is not allowed.} 
				and $G_{DS} = 0$ since all in-neighbors of $D$ are collected in $I$. 
				We now assume that the dource $D$ is not excited and we show that we can then not uniquely recover $G_{OD}$ and $G_{OI}$.  
				
				From the relationship $(I - G) T = I$, 
				we write down all equations in which $G_{OD}$ and $G_{OI}$ appear. 
				This yields the following.
				\begin{eqnarray}
				&&\hspace{-8mm}\begin{bmatrix}
				-G_{OD} & 1 & -G_{OI} & -G_{OS}
				\end{bmatrix}
				\left[
								\begin{array}{c c | c c}
												T_{DD} & T_{DO} & T_{DI} & T_{DS} \\
												T_{OD} & T_{OO} & T_{OI} & T_{OS} \\
												\hline
												T_{ID} & T_{IO} & T_{II} & T_{IS} \\
												T_{SD} & T_{SO} & T_{SI} & T_{SS} \\
								\end{array} 
				\right] \nonumber\\
				\nonumber\\
				&&\hspace{-8mm}= 
				\begin{bmatrix}
				0 & 1 & 0 & 0
				\end{bmatrix} \label{defeqn}
				\end{eqnarray}
				Since $D$ is a dource, it follows   that $G_{OD} \neq 0$ and that all elements of $G_{OI}$ are nonzero, whereas the elements of $G_{OS}$ can be zero or nonzero.
				Since it is assumed that $D$ is not excited, the first column of the $T$ matrix is unknown. 
				Disregarding this unknown first column of $T$,  
				we get the following equation. 
				\begin{align}
				&\begin{bmatrix}
				G_{OD} & G_{OI} 
				\end{bmatrix}
				\left[
				\begin{array}{c c  c}
												T_{DO} & T_{DI} & T_{DS} \\
												T_{IO} & T_{II} & T_{IS} \\
								\end{array} 
								\right]
				= \nonumber \\  
				&\begin{bmatrix}
								1 + T_{OO} - G_{OS} T_{SO} & T_{OI} - G_{OS} T_{SI} & T_{OS} - G_{OS} T_{SS}
				\end{bmatrix} \label{eq:mat2}
				\end{align}
				The question is whether or not $G_{OD}$ and $G_{OI}$  can be uniquely identified from equation (\ref{eq:mat2}), 
				even in the situation where all nodes other than $D$ are both excited and measured, i.e.~even if all $T_{XY}$ elements in (\ref{eq:mat2}) are known.

				Equation (\ref{eq:mat2}) has a unique solution for $[G_{OD}~ G_{OI}]$ 
				only if the matrix 
				\begin{align}
				\left[
				\begin{array}{c c  c}
												T_{DO} & T_{DI} & T_{DS} \\
												T_{IO} & T_{II} & T_{IS} \\
								\end{array} 
								\right]							
			\label{eq:TDIS}
				\end{align}
				has full row rank. 
				We show that this matrix is not full  rank. 
				Combining (\ref{eq:recursion2}) and (\ref{eq:recumirror}) in Lemma \ref{lem:recur} yields:				\begin{align}
								T_{ik} = T_{i, :} G_{: k} = G_{i, :} T_{:, k}
								\label{prodTG}
				.\end{align}
				where $T_{i, :}$ and $T_{:, k}$ denote the $i$-th row and $k$-th column of $T$, respectively. 
				Therefore we can write:
				\begin{align*}
								T_{DO} &= G_{D, :} T_{:, O} = 
								\begin{bmatrix} 0 & G_{DO} & G_{DI} & G_{DS} \end{bmatrix}
								\begin{bmatrix} T_{DO} \\ T_{OO} \\ T_{IO} \\ T_{SO} \end{bmatrix} \\
								&= 
								\begin{bmatrix} 0 & 0 & G_{DI} & 0 \end{bmatrix} T_{:, O} = G_{DI} T_{IO} \\
								\mbox{Similarly:}\\
								T_{DI} &= G_{D, :} T_{:, I} = G_{DI} T_{II}, \\
								T_{DS} &= G_{D, :} T_{:, S} = G_{DI} T_{I S}
				.\end{align*}
				Thus, the matrix in (\ref{eq:TDIS}) is not full rank, since the first row is a linear combination of the elements of the second block row. 
				As a result,  $G_{OD}$ and  $G_{OI}$ 
				can not be uniquely identified when the dource $D$  is not excited, which proves the result for dources, i.e.~condition 2) of the Theorem. The result for dinks, i.e.~condition 3) is obtained by duality, and the proof is therefore omitted.
\end{proof}
These new necessary conditions for identifiability of a dynamic network are both surprising and practically important. The definition of dources and dinks are based on the local topology around each node of the network.  
In the search for a valid EMP and, possibly, a minimal EMP, the first task is therefore to locate the dources and dinks in the network and to enforce, in the EMP, an excitation at each dource and a measurement at each dink; in addition, of course to the obvious excitation requirement for all sources and the measurement requirement for all sinks. An algorithm that detects all dources and dinks in the network from its matrix $G$ has been developed by Eduardo Mapurunga \cite{mapurunga_dources_dinks_2022}.

\section{Only sources and dources need to be excited}\label{sec:ifandonlyif}

In this section we move from the necessary condition on the excitation of sources and dources of 
Theorem~\ref{theo:dourcesanddinks} to a necessary and sufficient condition. We show that 
 sources and dources are the only type of nodes that need to be excited for the generic identifiability of a dynamic network, regardless of the excitation and measurement pattern (EMP) on all other nodes of the network. For reasons of brevity, this section is entirely devoted to the excitation condition for sources and dources. It is of course important to stress that the dual result holds for the measurement condition for sinks and dinks.

\begin{theorem}
				Consider a dynamic network with network matrix $G$ and let $D$ be a node of interest. 
				There exists at least one valid EMP in which node $D$ is measured but not excited if and only if node
				$D$ is neither a source nor a dource.
				\label{theo:iffNeitherSourceNorDource}
\end{theorem}
%
%
%
\begin{proof}
				The necessity of excitation requirement follows directly from Theorem \ref{theo:dourcesanddinks}. We thus need to prove that
				if node $D$ is measured but not excited, and if it is  neither a source nor a dource, then
				$G$ is generically identifiable under the assumptions of the Theorem.
%
				
				We define the following partition of the nodes of the network, with respect to the node $D$ of interest. This partition is different from the one used to prove Theorem~\ref{theo:dourcesanddinks}.			
				Let $O$ be the out-neighbors of $D$ with dimension $k_O$, 
				 let $I$ be the in-neighbors of $D$ that are not in $O$ (i.e.~$I \triangleq \mathcal{N}^{-}_D \setminus \{\mathcal{N}_D^+ \bigcap \mathcal{N}^{-}_D\}$) with dimension $k_I$, and let $S$ be the remaining nodes with dimension $k_S$.  
				Thus $G$ is partitioned as in (\ref{eq:Gpartition1}), but with different definitions for $D, O, I$ and $S$ than in Theorem~\ref{theo:dourcesanddinks}.

				Now define an EMP in which $D$ is measured but not excited, all sources are excited and all sinks are measured, and all other nodes are both excited and measured. We  show that if $D$ is not a source or a dource, then, even if $T_{:, D}$ is not available (because node $D$ is not excited), we can identify all transfer functions from the network. 
				
				It follows from our definition of $O$, $I$ and $S$ that $G_{DS}=0, G_{ID}=0$ and $G_{SD}=0$. Assuming that  the first column of $T$ is unknown, the  remaining modules of $G$ must then be identified from the relationship $ (I - G) T = I$ where the column $T_{:, D}$ has been disregarded:
				\begin{align}
								-
							  \begin{bmatrix} 
												-1       & G_{D O} & G_{D I} & 0 \\
												G_{O D} & G_{O O} -I & G_{O I} & G_{O S} \\
												0 & G_{I O} & G_{II} - I & G_{IS} \\
												0 & G_{S O} & G_{S I} & G_{SS} - I
								\end{bmatrix} 
								\\
								\times \begin{bmatrix}
												T_{D O} & T_{D I} & T_{D S} \\
												T_{O O} & T_{O I} & T_{O S} \\
												T_{I O} & T_{I I} & T_{I S} \\
												T_{S O} & T_{S I} & T_{S S} \\
								\end{bmatrix} 
								= 
								\begin{bmatrix}
												0 & 0 & 0 \\
												I & 0 & 0\\
												0 & I & 0 \\
												0 & 0 & I
								\end{bmatrix} 
								\label{GTequation}
				\end{align}
				We now define the  matrices $T_A, T_D, T_O, T_I, T_S$ and $T_B$ as follows.
				\begin{align}
							  	T_A \bydef 
								\begin{bmatrix}
												T_{D O} & T_{D I} & T_{D S} \\
												T_{O O} & T_{O I} & T_{O S} \\
												T_{I O} & T_{I I} & T_{I S} \\
												T_{S O} & T_{S I} & T_{S S} \\
								\end{bmatrix} 
								\bydef 
								\begin{bmatrix}
												T_D \\
												T_O\\
												T_I \\
												T_S
								\end{bmatrix} 
								\bydef
								\begin{bmatrix}
												T_D \\
												T_B
								\end{bmatrix} 
								\label{TATBdef}
				\end{align}
				Based on the assumptions, the matrix $T_A$, of size $n \times (n-1)$ is fully known. We note that the matrices $T_D, T_O, T_I, T_S$ have $n$ columns and, respectively, $1, k_O, k_I$ and $k_S$ rows, and that 
 $T_B$ is the $(n-1) \times (n-1)$ matrix composed of the last three block rows of $T_A$. We note that $T_B$ is nonsingular by the assumptions on the $G$  matrix.
 
 The first equation in (\ref{GTequation}) can be equivalently written as follows:
				\begin{align}
								T_{DO} &=  G_{DO} T_{O O} + G_{D I} T_{I O} \label{TD1}\\
								T_{DI} &= G_{DO} T_{OI} + G_{DI} T_{I I} \label{TD2} \\
								T_{DS} &= G_{DO} T_{OS} + G_{DI} T_{I S} 
				.
				\label{TDdependence}
				\end{align}
				This proves that the row $T_D$ is  a linear combination of the rows of $T_O$ and $T_I$

				Now we observe from (\ref{GTequation}) that the identification of the elements of the first row, and of the third and fourth block rows of $G$ all depend on linear equations of the form
				\begin{align*}
					\begin{bmatrix}G_{UV}& G_{WX} & G_{YZ}\end{bmatrix}	T_B = C					
				,\end{align*}
				where $C$ is a known matrix with elements depending on $T_A$. Since $T_B$ has full rank, this shows that all these elements of $G$ can be identified from the known elements  of $T_A$. 
				
				It remains to show that the second block row of $G$ can also be identified from (\ref{GTequation}). Equivalently, we need to show that the following equation has a unique solution (Recall the definitions made in (\ref{TATBdef})).
				\begin{align}
							  \begin{bmatrix} 
												G_{O D} & G_{O O} & G_{O I} & G_{O S} 
								\end{bmatrix} 
								\begin{bmatrix}
												T_{D} \\
												T_{O } \\
												T_{I } \\
												T_{S } 
								\end{bmatrix} 
								= 
								\begin{bmatrix}
												T_{OO} - I& T_{OI} & T_{OS}
								\end{bmatrix} 
								\label{G2Tequation}
				\end{align}
				To show this, we will need to require that $D$ is not a source nor a dource. 
				Assume thus that $D$ is neither a source nor a dource. The following facts then hold:\\
				(i) All elements of $G_{DI}$, and possibly some elements of $G_{DO}$, are $\neq 0$. \\
				(ii) Since $D$ is not a dource, it follows that, for each node $o \in O$, at least one element of $G_{o, I}$ is zero, where $G_{o, I}$ is the vector of edges connecting all elements of $I$ to that node $o$.
				This property implies that, in the equation (\ref{G2Tequation}) corresponding to that node $o$, at least one element of $G_{o,I}$ need not be identified (since it is zero) and that, therefore, the corresponding row (or rows) of the matrix $T_{I}$ disappears from that equation. We show that the remaining submatrix of $T_A$ in (\ref{G2Tequation}) has full rank and that the corresponding row of $[G_{O D}~G_{O O}~ G_{O I}~ G_{O S}]$ can be identified. 
				
				We prove the result for the first equation of (\ref{G2Tequation}). Thus, let 
								$o_1 \in O$
					be the first element of $O$, and let $I= \{i_1, i_2, \ldots, i_{k_I}\}$. Using property (ii) above, we can assume without loss of generality that $G_{o_1, i_1}=0$. As a result, in the first equation of (\ref{G2Tequation}), the first row of $T_I$ disappears, leaving a $(n-1)\times (n-1)$ submatrix of $T_A$; call this matrix  ${\tilde T}_A$. We show that this submatrix ${\tilde T}_A$ has full rank, by showing that it has the same row span as $T_B$. 
					
To do this, we first define the $(n-1)\times (n-1)$ matrix ${\tilde T}_B$, which is obtained from $T_B$ by just reordering its rows as follows:
\begin{align}
							  	{\tilde T}_B \bydef 
								\begin{bmatrix}
												T_I\\
												T_O \\
												T_S
								\end{bmatrix} 
								\label{TBTildedef}
				\end{align}

				We  observe that ${\tilde T}_A$ has the same rows as $T_B$, except that row $T_{i_1,I}$ is replaced by  $T_D$, the first row of $T_A$. 
				With the notations defined in (\ref{TATBdef}), and on the basis of the expressions derived in (\ref{TDdependence}), we can write the following relations 
				\begin{align}
				T_{D}&= G_{DO}T_O + G_{DI}T_I\\
				&= G_{DO}T_O + G_{D, i_1}T_{i_1, I} + \sum_{j=i_2}^{i_{k_I}}G_{D,j}T_{j,I}\\
				&= \begin{bmatrix} G_{D,i_1} & G_{D,i_2:O}
				\end{bmatrix}
				 \tilde{T}_B
				\end{align}	
				with $G_{D, i_2:O} \triangleq \begin{bmatrix}  G_{D,i_2} & \cdots & G_{D, i_{k_I}} & G_{D O} & 0 \end{bmatrix}$.
As a result, one can write:
\begin{align}
\tilde{T}_A = 
\begin{bmatrix} 
G_{D,i_1} & G_{D,i_2:O} \\
0 & I_{n-2}
\end{bmatrix}  \tilde{T}_B
\end{align}
where $I_{n-2}$ is the identity matrix  of size $n-2$. Remembering that  $G_{D,i_1}\neq 0$ we see that  $\tilde{T}_A$ is the product of a nonsingular upper-triangular matrix and the matrix $\tilde{T}_B$, which has the same row space as $T_B$. Thus $\tilde{T}_A$ is nonsingular.						
				The same argument applies to all other equations of 	(\ref{G2Tequation}), i.e.~for all other nodes $o \in O$.
				\end{proof}		

The dual version of Theorem~\ref{theo:iffNeitherSourceNorDource}
for sinks and dinks is expressed as follows.
\begin{theorem}
				Consider a dynamic network with network matrix $G$ and let $D$ be a node of interest. 
				There exists at least one valid EMP in which node $D$ is excited but not measured if and only if node
				$D$ is neither a sink or a source.
				\label{theo:iffNeitherSinkNorDink}
\end{theorem}
\begin{proof}
The proof is just as long and complicated as that for Theorem~\ref{theo:iffNeitherSourceNorDource}, but since it is entirely parallel to it, it is omitted.
\end{proof}

The main message of Theorems~\ref{theo:iffNeitherSourceNorDource} and \ref{theo:iffNeitherSinkNorDink} is the following. Provided some node of interest in a network is neither a source nor a dource, one can design a valid EMP for which this node is not excited. Similarly, provided some node of interest in a network is neither a sink nor a dink, one can design a valid EMP for which this node is not measured. This, of course, has important practical implications for the design of EMP, for example when a node is difficult to excite or difficult to measure. 

\section{Understanding dources and dinks}\label{sec:understanding}

In this section, we first explain the intuition behind the requirement for the excitation of a dource, using a very simple example. We then present another example to illustrate the result of Theorem~\ref{theo:iffNeitherSourceNorDource}.

\subsection{Example A}
\begin{figure}[h]
				\centering
				\includegraphics[width=0.45\columnwidth]{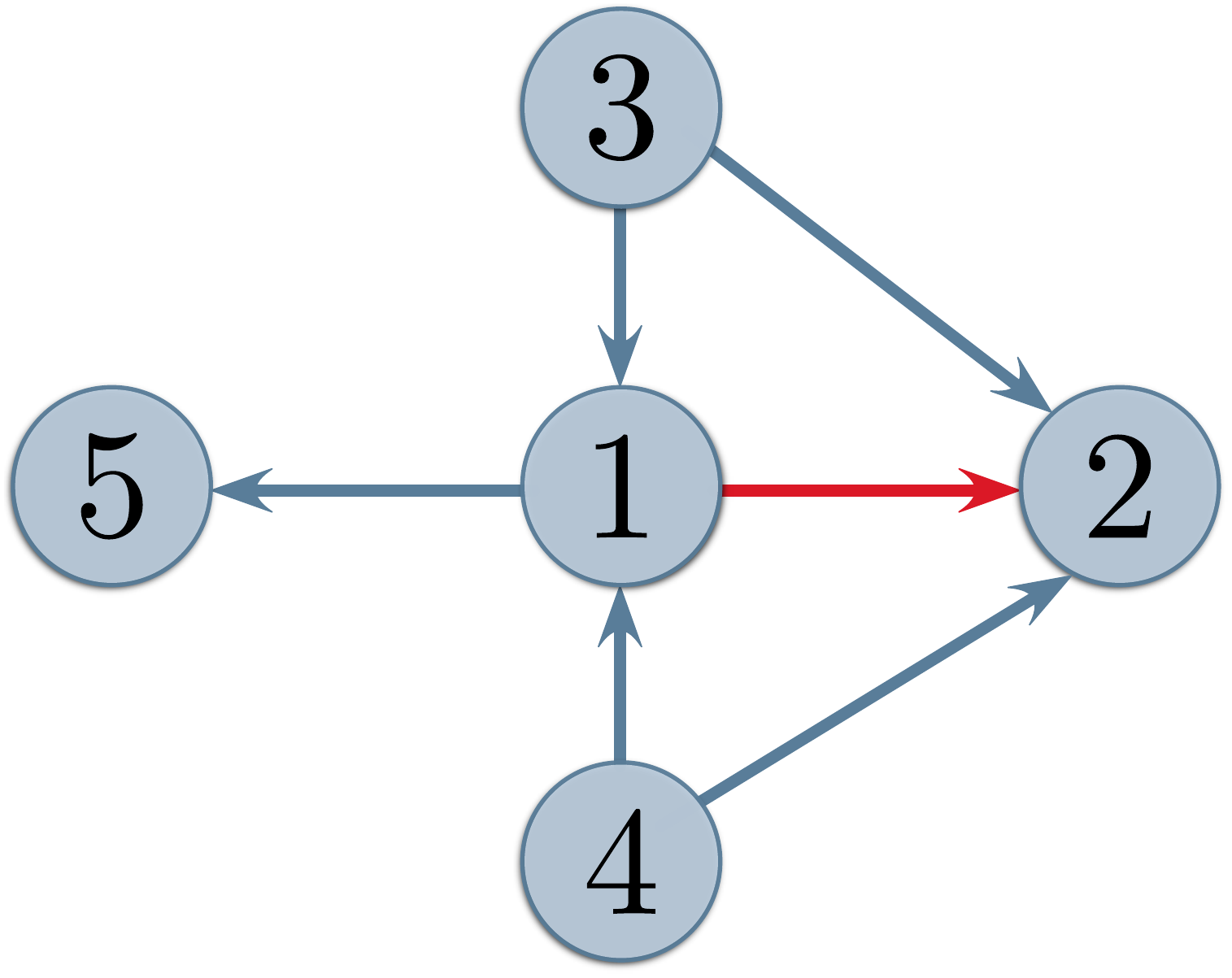}
				\caption{\normalsize Why does a dource require excitation?}
				\label{DD2DExample5}
\end{figure}

Consider the network of Figure~\ref{DD2DExample5}.
Its network matrix $G$ is:
\begin{equation}
G = \left[\begin{array}{ccccc}
0 & 0 & G_{13} & G_{14} & 0 \\
G_{21} & 0 & G_{23} & G_{24} & 0 \\
0 & 0 & 0 & 0 & 0  \\
0 & 0 & 0 & 0 & 0 \\
G_{51} & 0 & 0 & 0 & 0 
\end{array}\right]
\end{equation}
We observe that node $1$ is a dource, and we identify the sets defined in the proof of Theorem~\ref{theo:dourcesanddinks}
as $D=\{ 1\}$, $O=\{ 2\}$, $I=\{ 3, 4\}$, $S=\{ 5\}$.
The matrix in (\ref{eq:TDIS}) is given by 
\begin{eqnarray}
			&&\hspace{-5mm}	\left[
				\begin{array}{c c  c}
												T_{DO} & T_{DI} & T_{DS} \\
												T_{IO} & T_{II} & T_{IS} \\
								\end{array} 
								\right]							
			=
			\left[
				\begin{array}{c c  c}
												T_{12} & T_{1(3,4)} & T_{15} \\
												T_{(3,4)2} & T_{(3,4)(3,4)} & T_{(3,4)5} \\
								\end{array} 
								\right]							
			\nonumber
			\\
			&&=
			\left[
				\begin{array}{c c  c c}
												0 & G_{13} & G_{14} & 0 \\
												0 & 1 & 0 &  0 \\
												0 & 0 & 1 & 0 
								\end{array} 
								\right]							
				\end{eqnarray}
and it is clearly seen that there are only two nonzero equations to identify the three unknowns $G_{OD}= G_{21}$
and $G_{OI}=[G_{23}~G_{24}]$. Hence any EMP in which node $1$ is not excited is not valid; in other words, in all valid
EMPs node $1$ is excited. 

The intuition behind this result can be observed from the graph of Figure~\ref{DD2DExample5}. Nodes $3$ and $4$ are sources, and they must therefore be excited, while node $2$ must be measured because it is a sink. There are two parallel paths from the in-neighbors of the dource (i.e. nodes $3$ and $4$) to its out-neighbor node $2$. This makes the identification of the red edge $G_{21}$ impossible unless node $1$ is excited. The crucial condition for node $1$ to be a dource is that \emph{all} its neighbors must be connected to the out-neighbor $2$. If, on the other hand, the edge $(4,2)$ disappears, i.e. $G_{24}=0$, then node $1$ is no longer a dource and $G_{21}$ becomes identifiable using the excitation of node $4$, without exciting node $1$.

\subsection{Example B}

\begin{figure}[h!]
				\centering
				\includegraphics[width=0.45\columnwidth]{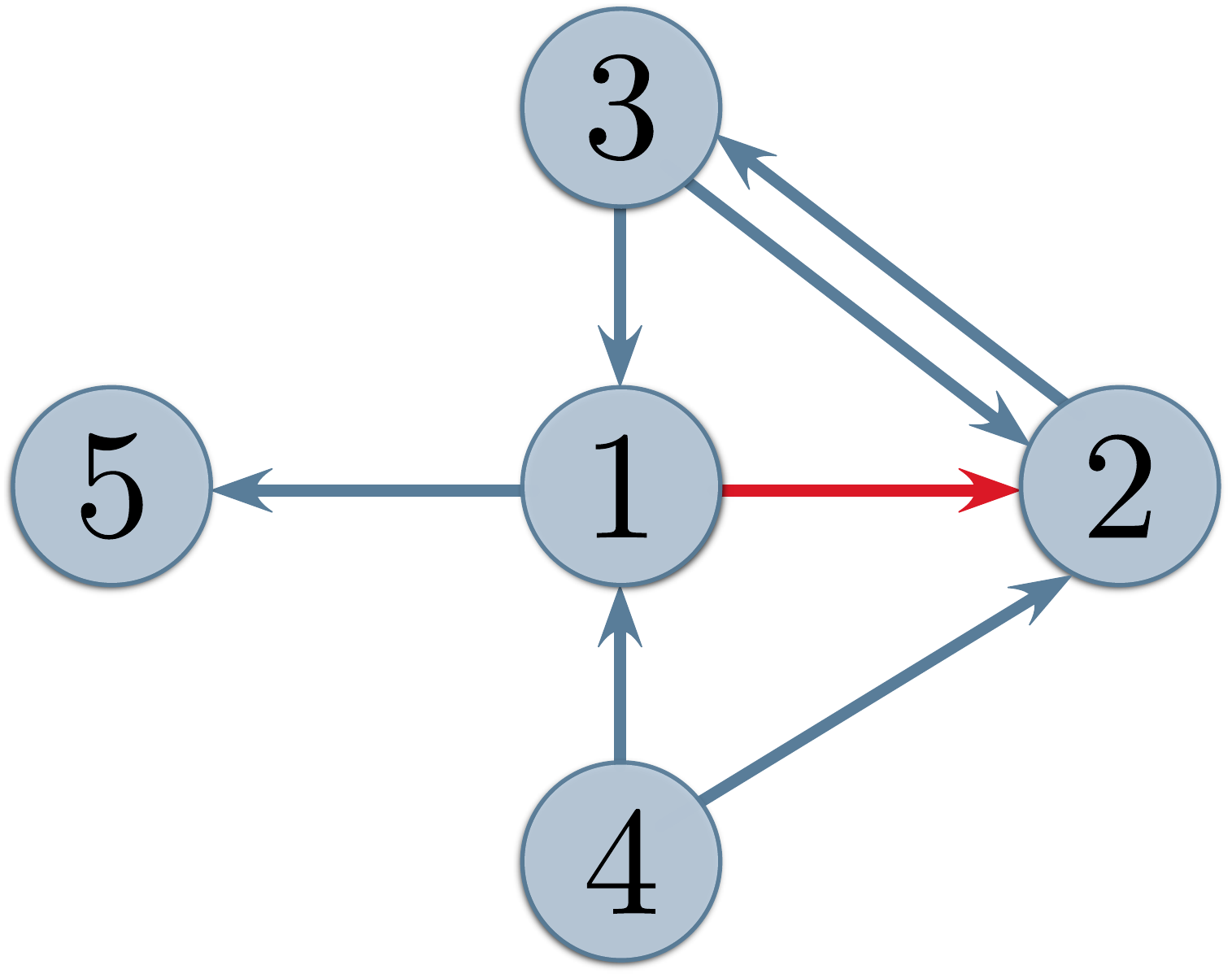}
				\caption{\normalsize Illustration of Theorem IV.1}
				\label{DD2DExample6}
\end{figure}

To illustrate the result of Theorem~\ref{theo:iffNeitherSourceNorDource} we use the example of Figure~\ref{DD2DExample6}. 

Its network matrix is given by
\begin{equation}
G = \left[\begin{array}{ccccc}
0 & 0 & G_{13} & G_{14} & 0 \\
G_{21} & 0 & G_{23} & G_{24} & 0 \\
0 & G_{32} & 0 & 0 & 0  \\
0 & 0 & 0 & 0 & 0 \\
G_{51} & 0 & 0 & 0 & 0 
\end{array}\right]
\end{equation}
\begin{figure*}[!t]
\normalsize
\setcounter{equation}{27}
\begin{equation}\label{T_exampleB}
T = \frac{1}{\Delta}
\left[\begin{array}{ccccc}
1-G_{23}G_{32} & G_{13}G_{32} & G_{13} & G_{13}G_{32}G_{24}-G_{14}G_{23}G_{32}+G_{14} & 0\\
G_{21} & 1 & G_{23}+G_{21}G_{13} & G_{21}G_{14}+G_{24} & 0 \\
G_{32}G_{21} & G_{32}  & 1 & G_{32}(G_{21}G_{14}+G_{24}) & 0 \\
0 & 0 & 0 & \Delta & 0 \\
G_{51}(1-G_{23}G_{32}) & G_{51}G_{13}G_{32} & G_{51}G_{13} & G_{51}(G_{13}G_{32}G_{24}-G_{14}G_{23}G_{32}+G_{14}) & \Delta
\end{array}\right]
\end{equation}
\setcounter{equation}{28}
\vspace*{4pt}
\end{figure*}
We observe that node $1$ is a dource. 
We identify the sets defined in the proof of Theorem~\ref{theo:iffNeitherSourceNorDource} as $D=\{ 1\}$, $O=\{ 2\}$, $I=\{ 3, 4\}$, $S=\{ 5\}$. The input-output matrix $T$ is given in (28) 
\ni with $\Delta = 1-G_{23}G_{32}-G_{21}G_{13}G_{32}$.
We establish the linear dependencies (\ref{TD1})-(\ref{TDdependence}), noting that $G_{DO}=0$.
\ni The relevant matrix is therefore  (\ref{eq:TDIS}), which  is given by
\begin{eqnarray*}
			&&\hspace{-8mm}	\left[
				\begin{array}{c c  c}
												T_{DO} & T_{DI} & T_{DS} \\
												T_{IO} & T_{II} & T_{IS} \\
								\end{array} 
								\right]							
			=
			\left[
				\begin{array}{c c  c}
												T_{12} & T_{1(3,4)} & T_{15} \\
												T_{(3,4)2} & T_{(3,4)(3,4)} & T_{(3,4)5} \\
								\end{array} 
								\right]							
			\nonumber
			\\
			&&\hspace{-8mm} =\!
			\frac{1}{\Delta}\!
			\left[
				\begin{array}{c  c  c c}
												G_{13}G_{32} & G_{13} & G_{13}G_{32}G_{24}\!-\!G_{14}G_{23}G_{32}\!+\!G_{14} & 0 \\
												G_{32} & 1 & G_{32}G_{14}G_{21}\!+\!G_{32}G_{24} &  0 \\
												0 & 0 & \Delta & 0 
								\end{array} 
								\!\! \right]							
				\end{eqnarray*}
\ni with $\Delta = 1-G_{23}G_{32}-G_{21}G_{13}G_{32}$. 
As stated in Theorem~\ref{theo:iffNeitherSourceNorDource}, the first row of this matrix is a linear combination of the other two: $l_1 = G_{13} l_2 + G_{14} l_3$
(here this is not evident by inspection, it requires some calculation). As a result, there are only two linearly independent equations to identify the three
unknowns $G_{OD}= G_{21}$ and $G_{OI}=[G_{23}~G_{24}]$. 
Hence any valid EMP requires that node $1$ be excited. 

Now, node $1$ is the only dource, and node $4$ is the only source. So, according to Theorem~\ref{theo:iffNeitherSourceNorDource},
the excitation of any other node is dispensable, in the sense that there is at least one valid EMP for which this node need not be excited. We illustrate the result with node $3$. 
Suppose that we do not excite it, so that the third column of $T$ in (\ref{T_exampleB})
is unknown. Identification of the network matrix must then rely on solving the set of equations 
\begin{align}
G \tilde T = \tilde T - \tilde I
\end{align}
where $\tilde T$ is the $(5 \times 4)$ matrix obtained by removing the third column of $T$ and $\tilde I$ is the $(5 \times 4)$  matrix obtained by removing the third column of the identity matrix $I_5$. This set of equations can be organized as four different 
systems of linear equations, one for each row of $G$:
\begin{eqnarray}
&&\hspace{-10mm}\left[\begin{array}{cc}
G_{13} & G_{14} 
\end{array}\right]
\left[\begin{array}{ccc}
T_{31} & T_{32} & T_{34}  \\
0 & 0 &  T_{44} 
\end{array}\right] \nonumber\\
\hspace{10mm}&&= 
\left[\begin{array}{ccc}
T_{11}-1 & T_{12}  & T_{14} 
\end{array}\right] \label{primeira}
\\
\nonumber \\
&&\hspace{-10mm}\left[\begin{array}{ccc}
G_{21} & G_{23} & G_{24} 
\end{array}\right]
\left[\begin{array}{ccc}
T_{11} & T_{12} &  T_{14}  \\
T_{31} & T_{32} &  T_{34} \\
0 & 0 &  T_{44} \\
\end{array}\right] \nonumber \\
\hspace{10mm}&& = 
\left[\begin{array}{ccc}
T_{21} & T_{22}-1 & T_{24} 
\end{array}\right] \label{segunda}
\\
&& \hspace{-10mm}G_{32} T_{21} = T_{31} \label{terceira}
\\
&&\hspace{-10mm}G_{51} T_{11}  =  T_{51}  \label{quarta}
\end{eqnarray}
where in all cases we have omitted the last column, which is zero. In  (\ref{primeira}) it is clear
from the matrix structure that there are two linearly independent equations for the two unknowns. In (\ref{segunda}),
it can be verified that the rank of the matrix is generically three by checking that 
$det (\left[\begin{array}{cc}
T_{11} & T_{12}   \\
T_{31} & T_{32}  \\
\end{array}\right]) = 
\frac{1}{\Delta}[G_{32}(1-G_{23}G_{32})-G_{21}G_{13}G_{32}^2)]= G_{32}\neq 0$
and that 
$T_{44} = 1$.
The other two equations (\ref{terceira}) and (\ref{quarta}) are both scalar,
 with $T_{21}, ~T_{11}\neq 0$. Notice that in both cases there are three scalar equations that
we could have used, and we picked the one corresponding to the first column of $T$.

Hence all four equations (\ref{primeira})-(\ref{quarta}) have a unique solution
and thus all the elements of the network matrix can be identified from knowledge of the first,
second and fourth columns, confirming that there is  no need
to excite node $3$. One valid EMP would thus be ${\cal B} = \{1, 2, 4 \}~~{\cal C} = \{1, 2, 3,  5\}$.

On the other hand, it is also easy  to see in this way that node $4$, which is a source, must be excited:
if the $4$-th column of $T$ is not known, then there are no coefficients multiplying $G_{14}$ or $G_{24}$ in any equations,
so these transfer functions cannot be identified.

\section{Conclusion}\label{sec:conclu}
This paper has put the spotlight on the essential r\^{o}le of two types of nodes whose existence in a network is easy to detect and depends on the local topology only, namely dources and dinks. Their r\^{o}le is essential in that they enforce constraints on the design of any valid Excitation and Measurement Pattern. Designing a valid EMP is the key to the identification of a dynamic network. With the results of this paper, we now know that without exciting all sources and all dources, and without measuring all sinks and all dinks, it is impossible to identify a dynamic network. The first task of any experiment designer is to detect all dources and dinks in the network. This can actually easily be achieved using the  algorithm designed for that purpose  by Eduardo Mapurunga; see \cite{mapurunga_dources_dinks_2022}. 

But in addition to this constraint, in the form of a necessary condition, this paper has also shown that if a specific node is not a source or a dource, the network can always be identified without exciting it. Dually, if a specific node is not a sink or a dink, the network can always be identified without measuring it.


\bibliographystyle{IEEETran}
\bibliography{StrucDyNetB2}

\end{document}